\documentclass[11pt]{article}
\usepackage[T1]{fontenc}
\usepackage{graphicx}
\usepackage{grffile}
\usepackage{longtable}
\usepackage{wrapfig}
\usepackage{rotating}
\usepackage[normalem]{ulem}
\usepackage{amsmath}
\usepackage{textcomp}
\usepackage{amssymb}
\usepackage{capt-of}
\usepackage{hyperref}
\usepackage[backend=biber, style=chicago-authordate, natbib=true, url=false, doi=false, uniquename=false]{biblatex}
\AtEveryBibitem{\clearlist{language}}
\AtEveryBibitem{\clearlist{ISSN}}
\usepackage{tabularx,amsthm}
\usepackage{dcolumn}
\usepackage[margin=2.5cm]{geometry}

\usepackage{setspace}
\usepackage{listings}
\usepackage{booktabs}
\usepackage[capposition=top]{floatrow}

\author{Joaquim Andrade, Pedro Cordeiro and Guilherme Lambais \thanks{U. Brasilia, Harvard Kennedy School and U. Brasilia, and U. Brasilia. Special thanks to Tiago Florido (in memoriam) for sharing code and initial conversations that greatly stimulated this project. We thank Isaiah Andrews, Jose Angelo Divino, Xavier Gabaix, and Michael Plagborg-Moller for very helpful comments and suggestions. We also thank Jordi Gali for sharing data and Argia Sbordone for help with the trend inflation measure in the first iterations of the paper. Part of this research was done when Lambais was a visiting scholar at the Department of Economics at Harvard University, he gratefully acknowledges Ed Glaeser and everyone there for their hospitality, and CNPq/CAPES funding.}}
\date{\today}
\title{Estimating a Behavioral New Keynesian Model}
\hypersetup{
 pdfauthor={Joaquim Andrade, Pedro Cordeiro and Guilherme Lambais \thanks{U. Brasilia, Harvard Kennedy School and U. Brasilia, and U. Brasilia. Special thanks to Tiago Florido (in memoriam) for sharing code and initial conversations that greatly stimulated this project. We thank Isaiah Andrews, Jose Angelo Divino, Xavier Gabaix, and Michael Plagborg-Moller for very helpful comments and suggestions. We also thank Jordi Gali for sharing data and Argia Sbordone for help with the trend inflation measure in the first iterations of the paper. Part of this research was done when Lambais was a visiting scholar at the Department of Economics at Harvard University, he gratefully acknowledges Ed Glaeser and everyone there for their hospitality, and CNPq/CAPES funding.}},
 pdftitle={Estimating a Behavioral New Keynesian Model},
 pdfkeywords={},
 pdfsubject={},
 pdfcreator={Emacs 26.1 (Org mode 9.1.14)}, 
 pdflang={English}}
\begin{document}

\maketitle
\begin{abstract}
This paper analyzes identification issues of a behavorial New Keynesian model and estimates it using likelihood-based and limited-information methods with identification-robust confidence sets. The model presents some of the same difficulties that exist in simple benchmark DSGE models, but the analytical solution is able to indicate in what conditions the cognitive discounting parameter (attention to the future) can be identified and the robust estimation methods is able to confirm its importance for explaining the proposed behavioral model.
\end{abstract}

\section{Introduction}
\label{sec:orgbca022f}

Optimization-based macroeconomic models usually rely on fully rational agents, but these models usually generate a number of paradoxes and taking these models to the data have not been totally successful; consumers and firms do not appear to be entirely forward looking. This paper estimates the macroeconomic model of \cite{Gabaix2019Behavioral} where agents are not fully rational. The behavioral mechanism in this model is fully microfounded and works in the form of a cognitive discounting. That is, as the agent simulate \(k\) steps into the future, the impact of her expectation is shrunk by a factor \(\bar{m}^k\) towards a simple benchmark, which can be the steady state of the economy. The parameter \(\bar{m} \in [0,1]\) captures this cognitive discounting, innovations far into the future get heavily discounted relative to the rational benchmark where \(\bar{m} = 1\). Another way of seeing this, is that the agent is globally patient in relation to the steady-state variables, yet is myopic with respect to deviations around the steady state, particularly if these deviations are in the remote future.

Aside from the theoretical issues that cognitive discounting brings to New Keynesian (NK) models and some impacts on monetary and fiscal policy that are already explored in \cite{Gabaix2019Behavioral}\footnote{New Keynesian models are mainly composed by the output and inflation equations and a monetary policy rule. The output equation generalizes the Euler equation for the whole economy and the inflation equation is a microfounded expectation-augmented Phillips curve. With fully rational agents, expectations in these curves are only forward looking, which generates a number of problems. The fully rational model generates the following predictions, which are continually contradicted by empirical facts: fiscal policy has no impact, depressions are moderate and bounded, equilibria is indeterminate at the zero lower bound, forward guidance by the central bank is very powerful, "price level targeting" is the optimal commitment policy, and the neo-Fisherian paradox (a rise in interest rates causes a rise in inflation). The model proposed by Gabaix solves all of these problems. This shows, as \cite{woodford2009} has argued, that despite the convergence in macroeconomic methodology in the last decades, important theoretical and empirical issues remain open. There is still little certainty on how to best specify an empirically adequate model of aggregate fluctuations.}, the estimation of macroeconomic models, either in the dynamic stochastic general equilibrium (DSGE) form or in the single equation form, has a very large literature on its own\footnote{See, for example, \cite{Fernandez-Villaverde2016Solution} for DSGE models and \cite{mavroeidisetal2014} for the NK Phillips Curve.}

Considering limited-information estimation of the New Keynesian Phillips Curve (NKPC), it is known that the purely forward-looking NKPC does not fit well aggregate U.S. inflation dynamics. There is still no consensus in the literature, but just to cite a few developments in response to this problem there is the "hybrid NKPC" (\cite{galigertler1999}), revisions to lower the size of the forward looking coefficient (\cite{ruddwhelan2005}), and inclusion of a trend inflation measure (\cite{cogleysbordone2008}). In addition, problems also arise in relation to the frequency of price re-optimization by firms (\textit{e.g.} \cite{eichenbaumfisher2007}). 

Likewise, numerous studies have found that the standard Euler equation model does not have a good fit to aggregate U.S consumption time series (see, \emph{e.g.} \cite{ascarietal2016} and \cite{fuhrerrudebusch2004}).\footnote{In a study about the Euler equation for consumption, \cite{havranek2015} conducts a meta-analysis of 169 published studies showing there is pervasive selective reporting of results and publication bias in this literature; exactly the problem we take head-on with the method implemented in this paper.} Extensions to the pure forward-looking Euler equation for output include habits (\cite{fuhrer2000}), hand-to-mouth consumers (\cite{bilbiiestraub2012}), and a hybrid version as well (\cite{mccallumnelson1998}).

Additionally, the estimation of full DSGE models saw remarkable advances in the past decades, however the estimation of large models can get complex and computationally demanding very quickly. While is out of the scope of this paper to discuss the literature of DSGE estimation in general, we are particularly interested in issues of weak identification that arise in both simple DSGE models and single-equation estimation derived from these DSGE models. In both strands of literature the issue of weak identification is abound, see, for example, \cite{Andrews2015Maximum} and \cite{Canova2009Back} for identification issues in DSGE models and \cite{mavroeidisetal2014}, \cite{ascarietal2016}, and \cite{Mavroeidis2010Monetary} for identification issues that arise in the estimation of the NK Phillips Curve, Euler equation, and monetary policy rules.

In this paper our contribution is, thus, twofold. First, we analytically solve a version of the model proposed by \cite{Gabaix2019Behavioral} to show where identification holds and it what conditions does it fail, and estimate the parameters of the system using robust maximum likelihood inference proposed in \textcite{Andrews2015Maximum}. Second, with a good understanding of what is driving identification we can then use less restrictive single-equation methods to estimate behavioral versions of the NK Phillips curve and the IS curve using two-step confidence sets proposed in \cite{andrews2018} that is also robust to identification failure.

The next section presents the behavorial NK model, section 3 solves the model and analyzes its identification, section 4 provides an initial likelihood-based estimation, section 5 presents the single-equation estimation method, section 6 the two-step robust confidence sets results and section 7 concludes.

\section{A Behavioral Macroeconomic Model}
\label{sec:org807104d}

We start with proposition 2.5 in \cite{Gabaix2019Behavioral}, a two-equation version of the Behavioral New Keynesian model, for the behavior of the output gap \(x_t\) and inflation \(\pi_t\):

\begin{equation}
\label{IS}
x_t = M\mathbb{E}_t[x_{t+1}] - \sigma (i_t - \mathbb{E}_t\pi_{t+1} - r_t^n) \;\; (IS\:curve),
\end{equation}

\begin{equation}
\label{NKPC}
\pi_t = \beta M^f \mathbb{E}_t[\pi_{t+1}] + \kappa x_t \;\; (Phillips\:curve),
\end{equation}
with \(i_t\) as the nominal interest rate, \(r_t^{n}\) is the natural interest rate, \(\sigma\) is the sensitivity of the output gap to the interest rate, \(\kappa\) is the sensitivity of the inflation to the output gap, and \(\beta\) is the pure rate of time preference. The equilibrium behavioral parameters \(M, M^f \in [0,1]\) are the aggregate-level attention parameters of consumers and firms, respectively, to macroeconomic outcomes:

\begin{equation}
M = \bar{m}, \;\; \sigma = \frac{1}{\gamma R},
\end{equation}

\begin{equation}
M^f = \bar{m} \left(\theta + \frac{1 - \beta \theta}{1 - \beta \theta \bar{m}} (1 - \theta)\right), \;\; \kappa = (\frac{1}{\theta} - 1)(1-\beta \theta)(\gamma + \phi)
\end{equation}
where \(\bar{m}\) is the myopia parameter, \(\theta\) is the survival rate of prices,  \(\gamma\) is the risk aversion, \(\sigma\) becomes the "effective" intertemporal elasticity of substitution, \(\phi\) is the inverse Frisch elasticity, and \(\kappa = (\frac{1}{\theta} - 1)(1-\beta \theta)(\gamma + \phi)\) is the slope obtained with fully rational firms. Firms are still fully attentive to the steady state, so they discount future profits at the rate \(R = \frac{1}{\beta}\). In the traditional benchmark model, \(\bar{m} = 1\), so that \(M = M^f = 1\).

The next section shows on what conditions does the identification of this model rest on and the procedures for identification-robust inference.

\section{Weak Identification in a Behavioral Dynamic Stochastic General Equilibrium Model}
\label{sec:orgbf7538d}

This section follows closely \cite{Andrews2015Maximum}. We first explore a dynamic stochastic general equilibrium (DSGE) version of the behavioral NK model by adding a monetary policy rule and exogenous technology and monetary policy shocks in addition to equations \ref{IS} and \ref{NKPC}, which gives the following system:

\begin{equation}
\begin{array}{rcl}
x_t & = & M\mathbb{E}_t[x_{t+1}] - \sigma (i_t - \mathbb{E}_t\pi_{t+1}) + \eta_{d,t} \\
\pi_t & = & \beta M^f \mathbb{E}_t[\pi_{t+1}] + \kappa x_t + \epsilon_{s,t}\\
i_t & = & \rho_i i_{t-1} + (1 - \rho_i)(\phi_\pi \pi_t + \phi_x x_t) + \eta_{m,t}
\end{array}
\end{equation}
where the unobserved exogenous shocks are generated by the law
\begin{equation}
\label{eq:org88761b0}
\begin{array}{rcl}
\eta_{d,t} & = & \rho_d \eta_{d,t-1} + \epsilon_{d,t}\\
\eta_{m,t} & = & \rho_m \eta_{m,t-1} + \epsilon_{m,t}\\
(\epsilon_{s,t}, \epsilon_{d,t}, \epsilon_{m,t})' & \sim & \text{i.i.d.}N(0,\Sigma)\\
\Sigma & = & \text{diag}(\sigma^2_s, \sigma^2_d, \sigma^2_m).
\end{array}
\end{equation}

We make several simplifying assumptions to be able to solve the model analytically. Specifically, assume that \(\rho_i=0\), \(\phi_x=0\), \(\phi_{\pi}=\frac{1}{\sigma}\), and \(\sigma^2_s=0\), so the model has \(\vartheta = (\beta, \theta, \bar{m}, \gamma, \phi, \rho_m, \rho_d, \sigma^2_m, \sigma^2_d)\), knowing that \(M^f\), \(\sigma\), and \(\kappa\) are functions of \(\bar{m}\), \(\beta\), \(\theta\), \(\gamma\), and \(\phi\). In Section \ref{sec:org0af572b} of the Appendix using these restrictions we obtain the following solution for the behavioral DSGE model
\begin{equation}
\label{eq:org8c224c1}
\begin{pmatrix}
x_t \\
\pi_t
\end{pmatrix}
=
\begin{pmatrix}
\dfrac{-\beta M^f \sigma}{\beta M^f + \sigma \kappa - \rho_m \bar{m}} & \dfrac{\beta M^f}{\beta M^f + \sigma \kappa - \rho_d \bar{m}} \\[2ex]
\dfrac{-\beta M^f \sigma \kappa}{(\beta M^f + \sigma \kappa - \rho_m \bar{m})(1 - \rho_m \beta M^f)}  & \dfrac{\beta M^f}{(\beta M^f + \sigma \kappa - \rho_d \bar{m})(1 - \rho_d \beta M^f)}
\end{pmatrix}
\begin{pmatrix}
\eta_{m,t} \\
\eta_{d,t}
\end{pmatrix}
\end{equation}

To analyze the identification of the model parameters, let

\begin{equation}
A_1(\vartheta) = \frac{-\beta M^f \sigma}{\beta M^f + \sigma \kappa - \rho_m \bar{m}} \;\;\;\text{and}\;\;\; A_2(\vartheta) = \frac{\beta M^f}{\beta M^f + \sigma \kappa - \rho_d \bar{m}},
\end{equation}
thus we can write each equation in the system \ref{eq:org8c224c1} for \(x_t\) and \(\pi_t\) as
\begin{equation}
\begin{array}{rcl}
x_t & = & A_1(\vartheta)\eta_{m,t} + A_2(\vartheta)\eta_{d,t} \\[2ex]
\pi_t & = & \dfrac{\kappa}{1-\rho_m \beta M^f} A_1(\vartheta) \eta_{m,t} + \dfrac{\kappa}{1-\rho_d \beta M^f} A_2(\vartheta) \eta_{d,t}.
\end{array}
\end{equation}

We can now express the autocovariances and cross-covariances of the series \(x_t\) and \(\pi_t\) using the two equations above and the law of motion in \ref{eq:org88761b0}. In particular, for \(x_t\) the autocovariances are

\begin{equation}
\begin{array}{lcl}
\text{Var}(x_t) & = & A_1(\vartheta)^2 \dfrac{\sigma^2_m}{1-\rho^2_m} + A_2(\vartheta)^2 \dfrac{\sigma^2_d}{1-\rho^2_m} \\[2ex]
\text{Cov}(x_t, x_{t-k}) & = & A_1(\vartheta)^2 \dfrac{\sigma^2_m \rho^k_m}{1-\rho^2_m} + A_2(\vartheta)^2 \dfrac{\sigma^2_d \rho^k_d}{1-\rho^2_m}
\end{array}
\end{equation}
from which we can identify \(\rho_d \neq \rho_m\), \(A_1(\vartheta)^2 \sigma^2_m\), \(A_2(\vartheta)^2 \sigma^2_d\). Additionally, the expression for the cross-covariance structure of the processes \(x_t\) and \(\pi_t\) is

\begin{equation}
\begin{array}{lcl}
\text{Cov}(x_t, \pi_t) & = & A_1(\vartheta)^2 \dfrac{\sigma^2_m}{1-\rho^2_m} \dfrac{\kappa}{1-\rho_m \beta M^f} + A_2(\vartheta)^2 \frac{\sigma^2_d}{1-\rho^2_a} \dfrac{\kappa}{1-\rho_d \beta M^f} \\[2ex]
\text{Cov}(x_t, \pi_{t-k}) & = & A_1(\vartheta)^2 \dfrac{\sigma^2_m \rho^k_m}{1-\rho^2_m} \dfrac{\kappa}{1-\rho_m \beta M^f} + A_2(\vartheta)^2 \dfrac{\sigma^2_d \rho^k_d}{1-\rho^2_a} \dfrac{\kappa}{1-\rho_d \beta M^f}.
\end{array}
\end{equation}
from this structure we can identify \(A_1(\vartheta)^2 \sigma^2_m \frac{\kappa}{1 - \rho_m \beta M^f}\) and \(A_2(\vartheta)^2 \sigma^2_d \frac{\kappa}{1 - \rho_d \beta M^f}\). 

Thus, in all from the autocovariance structure of processes \(x_t\) and \(\pi_t\), if \(0 < \beta < 1\), \(0 < \bar{m} < 1\), \(0 < \theta < 1\), \(0 < \rho_m < 1\), \(0 < \rho_d <1\),  \(\kappa >0\), \(\sigma^2_m > 0\), and \(\sigma^2_d > 0\) we can identify six quantities
\begin{equation*}
\rho_m,\;\; \rho_d,\;\; A_1(\vartheta)^2 \sigma^2_m, A_2(\vartheta)^2 \sigma^2_d,\;\; A_1(\vartheta)^2 \sigma^2_m \frac{\kappa}{1 - \rho_m \beta M^f},\;\; A_2(\vartheta)^2 \sigma^2_d \frac{\kappa}{1 - \rho_d \beta M^f}.
\end{equation*}
Looking at the last four quantities we can see that \(\frac{\kappa}{1-\rho_m \beta M^f}\) and \(\frac{\kappa}{1-\rho_d \beta M^f}\) are identified, thus \(\frac{1-\rho_m \beta M^f}{1-\rho_d \beta M^f}\) is identified. Since \(\rho_d\) and \(\rho_m\) are part of the six quantities initially identified, we have that the product \(\beta M^f\) is identified as well. The parameter \(M^f\) is equal to \(\bar{m} \left(\theta + \frac{1 - \beta \theta}{1 - \beta \theta \bar{m}} (1 - \theta)\right)\), thus if we fix a value for \(\beta\) and \(\theta\), which is common in the literature, then \(\bar{m}\) is identified. Furthermore, this implies that \(\kappa\) is identified. Since \(\kappa\) is equal to \((\frac{1}{\theta} - 1)(1-\beta \theta)(\gamma + \phi)\), if we fix a value for \(\phi\), then \(\gamma\) is also identified. Now since \(\sigma\) is equal to \(\frac{\beta}{\gamma}\), \(\sigma\) is identified as well. With these quantities identified so far, it implies that \(\sigma^2_m\) and \(\sigma^2_d\) are identified. To sum up, we have three degrees of underidentification - nine structural parameters but only six identified quantities -, thus we have to fix three parameters to identify the other six.

If \(\rho_d = \rho_m\) the situation is different. If \(\rho_d = \rho_m\) then the series for \(x_t\) and \(\pi_t\) becomes

\begin{equation}
\begin{array}{rcl}
x_t & = & \dfrac{\beta M^f}{\beta M^f + \sigma \kappa - \rho_{m,d} \bar{m}} (\eta_{d,t} - \sigma \eta_{m,t}) \\[2ex]
\pi_t & = & \dfrac{\beta M^f}{(\beta M^f + \sigma \kappa - \rho_{m,d} \bar{m})(1 - \rho_{m,d} \beta M^f)} (\eta_{d,t} - \sigma \eta_{m,t}) = \dfrac{\kappa}{1-\rho_{m,d} \beta M^f} x_t.
\end{array}
\end{equation}
which are linearly dependent AR(1) processes with autoregressive root \(\rho_m = \rho_d\). From this system with can only identify four quantities: the autogressive parameter \(\rho_m = \rho_d\), the variance of \(x_t\), and the ratio \(x_t / \pi_t\),

\begin{equation*}
\rho_m = \rho_d,\;\; \frac{\beta M^f}{\beta M^f + \sigma \kappa - \rho_{m,d} \bar{m}}  \sqrt{\sigma^2 \sigma^2_m + \sigma^2_d},\;\; \frac{\kappa}{1-\rho_{m,d} \beta M^f}.
\end{equation*}

Hence, we now have two extra degrees of underidentification. More importantly, even if \(\rho_d \neq \rho_m\), as the difference \(\rho_d - \rho_m\) approaches zero there is a difficulty in making reliable statistical inferences. Following the example in \cite{Andrews2015Maximum}, take the Wald statistic \(W\) for testing the true hypothesis \(H_0:\vartheta = \vartheta_0\). Under usual asymptotic theory of maximum likelihood, if \(\rho_m \neq \rho_d\) then as the sample size \(T\) increases to infinity, the statistic \(W\) converges in distribution to \(\chi^2_9\) under \(H_0\). However, if \(\rho_m = \rho_d\) this convergence breaks down in the limit distribution of \(W\). The distribution of \(W\) experiences a discontinuity at \(\rho_m = \rho_d\), which implies that the convergence to \(\chi^2\) is not uniform in the parameter \(\rho_d - \rho_m\) in the neighborhood of zero. 

The consequences can be quite severe in distorting the size of the test. For example, \cite{Andrews2014Weak} documents for simple DSGE model that if \(\rho_m - \rho_d = 0.05\), then the size of a \(5\) percent Wald test is actually \(88.9\%\) and even for a large difference of \(\rho_d - \rho_m = 0.7\) the size of the test is \(9.8 \%\), that is, instead of falsely rejecting \(H_0\) only the standard \(5\%\) of the times, one would be falsely rejecting \(H_0\) between approximately \(90\%\) and \(10\%\) most of the times.

\section{Maximum Likelihood Inference with Robust Confidence Sets}
\label{sec:org6d3039f}

From the solution of the DSGE model in equation \ref{eq:org8c224c1} we have a space-state representation of the system and we readily apply the maximum likelihood method. The estimation in this section is made using the complete model, but to illustrate the method we proceed with the simplified solution, which is rearranged as

\begin{equation}
Y_t = 
\begin{pmatrix}
x_t \\
\pi_t
\end{pmatrix}
= C(\vartheta)
\begin{pmatrix}
\eta_{m,t} \\
\eta_{d,t}
\end{pmatrix}
= C(\vartheta)U_t
\end{equation}

and

\begin{equation}
U_t = \Lambda U_{t-1} + \epsilon_t,\;\;\; \Lambda=\begin{pmatrix} \rho_m & 0 \\
0 & \rho_d
\end{pmatrix}
\text{and}\;\epsilon_t \sim N(0,\Sigma).
\end{equation}
The log likelihood of the state-space system is:

\begin{equation}
\begin{array}{lcl}
\ell_T(\vartheta) & = & \text{const} \\[2ex]
{} & {} & - \frac{1}{2}\sum^T_{t=1}(C^{-1}(\vartheta)Y_t - \Lambda C^{-1}(\vartheta)Y_{t-1})' \Sigma^{-1} (C^{-1}(\vartheta)Y_t - \Lambda C^{-1}(\vartheta)Y_{t-1}) \\[2ex]
{} & {} & -\frac{T}{2}\text{log}|\Sigma|-T\text{log}|C(\vartheta)|.
\end{array}
\end{equation}

The full model has as endogenous observed series, \(i_t\), the Effective Federal Funds Rate, in addition to \(x_t\), the output gap, and \(\pi_t\), the inflation rate. Data for these three series are taken from the Federal Reserve Bank of St. Louis' FRED database for the period 1962:Q2 to 2016:Q4 and detrended. 

Table \ref{org9ed47be} reports maximum likelihood estimates using Chris Sims minimization routine that employs a quasi-Newton method with BFGS updates of the estimated inverse hessian for the structural model parameters \(\bar m\), \(\gamma\), \(\phi_\pi\), and \(\phi_x\) and for the shock parameters \(\rho_i\), \(\rho_d\), \(\rho_m\), \(\sigma^2_d\), \(\sigma^2_s\), and \(\sigma^2_m\) restricting \(\beta=0.99\), \(\theta=0.875\), and \(\phi=1\).

The estimate for \(\bar m\) is \(0.67\) with standard deviation of \(0.07\) and highly significant. However, since \(\rho_d = 0.95\) and \(\rho_m = 0.88\) making the difference between them less than \(0.2\) and thus highly susceptible to size distortions in the tests.

In this context, \textcite{Andrews2015Maximum} presents a robust test to generate confidence intervals for the model parameters. This approach uses a improved version of the \(LM\) test that is robust to weak identification. That is, the derivation of the asymptotic distribution of the \(LM\) statistics does not use any assumption about the strength of identification. The \(LM\) statistics is calculated using the score function and the (theoretical) Fisher information that can be calculated either using the negative Hessian of the log likelihood or the quadratic variation of the score. Both deliver unbiased estimates of the (theoretical) Fisher information for the whole sample and differ only in computational implementation. We use a version of the test with the quadratic variation of the score that is equivalent to

\begin{equation}
J_T(\vartheta) = [S_T(\vartheta)] = \sum^T_{t=1}s_{T,t}(\vartheta)s'_{T,t}(\vartheta),
\end{equation}
where \(s_{T,t}(\vartheta)\) is the increment of the score function \(S_T(\vartheta) = S_{T,T}(\vartheta) = \frac{\partial}{\partial \vartheta'} \ell_T(X_T, \vartheta)\) and \(X_T\) the data available at time \(T\). Under conditions expressed in \textcite{Andrews2015Maximum},

\begin{equation}
LM_o(\vartheta_0) = S_T(\vartheta_0)J_T(\vartheta_0)^{-1}S_T(\vartheta_0) \Rightarrow \chi^2_k
\end{equation}
with \(k = dim(\vartheta_0)\).

Then to calculate a \(95\%\) \(LM_o\) confidence set for the parameter \(\vartheta\) such that \(H_0: \vartheta = \vartheta_0\) is not rejected by an \(LM_o\) test with size \(5\%\) we first divide the parameters in incremental groups:

\begin{enumerate}
\item \(\vartheta = (\bar{m}, \gamma, \phi_\pi, \phi_x, \rho_i)\)
\item \(\vartheta = (\bar{m}, \gamma, \phi_\pi, \phi_x, \rho_i, \rho_d)\)
\item \(\vartheta = (\bar{m}, \gamma, \phi_\pi, \phi_x, \rho_i, \rho_m)\)
\item \(\vartheta = (\bar{m}, \gamma, \phi_\pi, \phi_x, \rho_i, \sigma_d)\)
\item \(\vartheta = (\bar{m}, \gamma, \phi_\pi, \phi_x, \rho_i, \sigma_s)\)
\item \(\vartheta = (\bar{m}, \gamma, \phi_\pi, \phi_x, \rho_i, \sigma_m)\).
\end{enumerate}

\textcite{Andrews2015Maximum} shows how doing composite hypotheses controls the size distortion of the test and so we follow the same strategy here. Next, we draw samples from the model with parameters calibrated to ML estimates obtained in Table \ref{org9ed47be}. The model is point identified at these values. We generate samples with 400 observations and discard the first and last 100. Using this random draw, we treat it as a sample and test each group of parameters with \(10^4\) uniform draws at random over the parameter space \(\vartheta\) delineated in items 1-6 and collect all values that the corresponding hypothesis \(H_0: \vartheta = \vartheta_0\) are not rejected. Then by projecting the five-dimensional convex set obtained in (1) on the subspace corresponding to each parameter separately, we obtain one-dimensional confidence sets for \((\bar{m}, \gamma, \phi_\pi, \phi_x, \rho_i)\). To obtain a confidence set for the remaining parameters we project the corresponding six-dimensional sets (2)-(6) on the subspace of the parameter of interest.

Table \ref{orgc0f1c14} presents the results for this procedure. The confidence intervals are wide but in most cases they exclude a wide range of values and in some cases they cover only small part of the parameter space, thus generating useful information. The confidence interval for \(\bar{m}\) is \([0.013, 0.645]\).

\section{Single-equation Estimation and Identification-robust Confidence Sets}
\label{sec:org670343b}

In this section we relax some assumptions about the model and the data generating function. Adding unrestricted innovations to equations 1 and 2, \(u_i\) and \(e_i\), which can represent unobserved cost-push shocks (either to the markup or input prices) in the case of the Phillips curve and an aggregate demand shock in the case of the IS curve, we obtain the "semi-structural" version of the model.

In this case we can demonstrate how a Generalized Instrumental Variables (GIV) approach with a Generalized Method of Moments (GMM) estimator can be valid.\footnote{The GIV approach was first proposed for the estimation of rational expectation models by \cite{mccallum1976} and then \cite{hansen1982} in the context of estimation of Euler equations. More recently it has been proposed for the estimation of the NKPC by \cite{roberts1995} and \cite{galigertler1999} (see also \cite{mavroeidisetal2014}).} Conditions for identification hinge on exclusion restrictions implied by excluding lags of the model and using them as instruments. The most common implementation of the GIV procedure substitutes the rational expectation by its realization. With this substitution and the addition of the idiosyncratic shocks, equations 1 and 2 become

\begin{equation}
\pi_t = \beta M^f \pi_{t+1} + \kappa x_t + \underbrace{u_t + \underbrace{\beta M^f[\pi_{t+1} - \mathbb{E}(\pi_{t+1})]}_{\text{$\pi$ forecast error}}}_{\text{$\tilde{u}$}}
\end{equation}

\begin{equation}
x_t =  Mx_{t+1} -  \sigma (i_t - \pi_{t+1} - r_t^n) + \underbrace{e_t + \underbrace{M [x_{t+1} - \mathbb{E}(x_{t+1})]}_{\text{$x$ forecast error}} + \underbrace{\sigma [\pi_{t+1} - \mathbb{E}(\pi_{t+1})]}_{\text{$\pi$ forecast error}}}_{\text{$\tilde{e}$}}.
\end{equation}

Let \(\vartheta^1 = (\bar{m},\theta)\) and \(\vartheta^2 = (\bar{m},\beta)\) , and define the "residual" function of both equations

\begin{equation}
h_t^1(\vartheta^1) = \pi_t -  \beta M^f \pi_{t+1} - \kappa x_t
\end{equation}
and

\begin{equation}
h_t^2(\vartheta^2) = x_t - M\mathbb{E}_t[x_{t+1}] +  \sigma (i_t - \mathbb{E}_t\pi_{t+1} - r_t^n)
\end{equation}
with the assumption that there exists two vectors of valid instruments, \(Z_t^1\) and \(Z_t^2\), such that
\begin{equation}
\label{eq:org6496e84}
\mathbb{E}[Z_t^i h_t^i(\vartheta^i)] = 0 \;\; \forall i=1,2
\end{equation}
holds at the true parameter value \(\vartheta^i = \vartheta_0^i \;\; \forall i=1,2\).

The efficient GMM estimator is based on the sample moments \(f_T(\vartheta^i) = T^{-1} \sum_{t=1}^{T} Z_t^i h_t^i(\vartheta^i)\) and a heteroskedasticity and autocorrelation (HAC) consistent estimator of their variance,  because of possible autocorrelation of \(\tilde{u}\) and \(\tilde{e}\) due to the presence of forecast errors. Specifically, we use the \cite{neweywest1987} covariance estimator with four lags. Given \(f_T(\vartheta^i)\), the estimator wants to minimize the GMM objective function
\begin{equation}
\label{eq:org8a782df}
S_T(\vartheta^i, \bar{\vartheta^i}) = f_T(\vartheta^i)'W_T(\bar{\vartheta^i})f_T(\vartheta^i)
\end{equation}
with respect to \(\vartheta^i\), where \(W_T\) is weighting matrix. Setting \(\bar{\vartheta^i} = \vartheta^i\) and evaluating \(W_T(\vartheta^i)\) at the same parameters as \(f_T(\vartheta^i\)) gives us the continuous updating estimator (\cite{hansenetal1996}).\footnote{Two-step GMM and continuous updating GMM (CUGMM) are asymptotically equivalent under strong identification, but CUGMM has some advantages under weak identification (\cite{stocketal2002}), also it is the preferred estimator in Andrew's method presented ahead.}

A common identifying assumption in the literature for both the Phillips and the IS curves is that both cost-push and aggregate demand shocks satisfy \(\mathbb{E}_{t-1}(u_t) = 0\) and \(\mathbb{E}_{t-1}(e_t) = 0\). Given the rational expectations assumption and the law of iterated expectations, the identifying assumption yields \(\mathbb{E}_{t-1}(\tilde{u_t}) = 0\) and \(\mathbb{E}_{t-1}(\tilde{e_t}) = 0\). Thus, we can have unconditional moment restriction for the form of equation \ref{eq:org6496e84}  with \(Z_t^i = Y_{t-1}^i\), for any vectors of predetermined variables. Any vector of variables \(Y\) known at time \(t-1\) can be used as instruments and implementations of GIV will differ in these choices. In this paper we take a novel approach in the sense that we don't pretest or screen for sets of instruments prior to estimation.

To implement the estimation we use the fully structural version of the model guided by the identification analysis and restrictions already imposed on the DSGE model. Particularly for the single equation setting, because of the difficulty to forecast inflation and the output gap, weak instruments is a pervasive problem that threatens the validity of structural inference under any identification approach (\cite{mavroeidis2004}). In addition, in a setting with weak instruments one would want a identification-robust method to avoid selective reporting\footnote{An interesting thought experiment put forth by I. Andrews is the following: imagine a world in which no instruments have any identifying power whatsoever, then by pure chance in a linear application we will sometimes observe a large value of the first-stage F statistic, but still any confidence set reported based on this screening will be invalid.} and control coverage distortions. \textit{Identification robust} is understood in the sense that if point identification fails the robust confidence set still covers the true parameter value. In other words, the confidence sets are uniformly asymptotically valid even when we allow for near or complete identification failure. Thus we need to derive a test statistic whose distribution under the null is insensitive to weak identification.

Fortunately, \cite{andrews2018} has already shown that there are statistics with good properties for this setting. To apply his method we proceed in two steps. Represent the outcome of the first step using a identification category selection (ICS) statistic \(\phi_{ICS} \in {0,1}\). Where \(\phi\) is some test statistic and \(\phi_{ICS}=1\) indicates evidence of weak identification and \(\phi_{ICS}=0\) of strong identification. In the second step we use: \(CS_{N}\) if identification seems strong and \(CS_R\) if identification seems weak. We can write two-step confidence sets as

\begin{equation}
  CS_{2S}=\begin{cases}
    CS_{N} & \text{if $\phi_{ICS}=0$}\\
    CS_R & \text{if $\phi_{ICS}=1$}
  \end{cases}
\end{equation}

and we are interested in the coverage of this two-step confidence set \(Pr_{\vartheta_0^i}\)\{\(\vartheta_0^i \in CS_{2S}\)\}. And we will assume \(CS_{N}\) has coverage of at least \(1-\alpha\) under strong identification and \(CS_R\) has coverage at least \(1-\alpha\) under both weak and strong identification. We define the maximal coverage distortion for \(CS_{2S}\) as the smallest \(\Gamma\) such that \(Pr_{\vartheta_0^i}\)\{\(\vartheta_0^i \in CS_{2S}\)\}\(\geq 1 - \alpha -\Gamma\). 

This procedure has mainly two elements: controlled coverage distortion of the test and test inversion. Controlled coverage distortion works by using a linear combination of the K and S statistics that have be shown to have good properties in these settings (\cite{andrews2016}). The robust confidence thus has a statistic produced by a linear combination of \(K\) and \(S\) statistics which derivation does not depend on the strength of identification, with this we have a coverage of \(1-\alpha\) for sure. Then we can see how much do we need to distort the test size for the confidence interval to fit in the non robust confidence set which test statistic is a conventional \(W\) statistic and has coverage \(1-\alpha-\hat{\Gamma}\), where \(\hat{\Gamma}\) is defined in this process as shown in Figure \ref{org16a9b50}. In Appendix \ref{sec:org1a4b023} we present the test and algorithm in more detail.

We present the results in this section using \(\alpha=0.05\) and \(\Gamma_{min}=0.05\), in the Appendix \ref{sec:orgbc247d4} we present results for \(\alpha=0.10\). The results are broadly similar but in some cases we can get smaller confidence sets.

For the behavioral IS curve the parameter grid used was \(\vartheta = (\bar{m}, \gamma) \in \Theta_D = {0.01, 0.02, ..., 0.99} \times {0.01, 0.02, ..., 10}\). Table \ref{orgb5f0b9c} presents the results illustrated in Figure \ref{orgde53aaf}. The CUGMM point estimates are \(\bar{m}=0.9029\) and \(\gamma = 2.281\) and the distortion cutoff \(\hat \Gamma\) is \(0.068\) for the entire set. This means that for one to believe in the non-robust set one has to be willing to add \(6.8\%\) on top of the original test size of \(5\%\). Looking individually, the distortion cutoff are a bit lower at the minimum \(5\%\) for both \(\bar m\) and \(\gamma\). The robust set is valid at the \(5\%\) level. If one is willing to make the trade-off between uncertainty and a tighter confidence set, then the predicted value of \(\bar m\) lies between \(0.80\) and \(1.00\), while for \(\gamma\) it is between \(1.07\) and \(3.49\).

Figure \ref{orgbeacf7f} illustrates the results for the behavioral NKPC estimation. The parameter grid is the same. The CUGMM point estimates are \(\bar{m}=0.393\) and \(\gamma=7.944\) and the distortion cutoff \(\hat \Gamma\) is \(14\%\) for the entire set. Table \ref{org6b6ad4f} details the results for each parameter. The distortion cutoff is a lower at \(9.93\%\) each. More importantly, there is an upper bound for \(\bar m\) at \(0.95\) in the robust case and at \(0.84\) in the non-robust case and lower bound of 0.07 in the non-robust case and 0.14 in the robust case.

\section{Conclusion}
\label{sec:org4afa8fa}

In this paper we analyzed the identification issues of a behavioral New Keynesian model and estimated it with likelihood-based and limited-information methods with identification-robust confidence sets. As a result we are able to, in the first place, validate to a certain degree \cite{Gabaix2019Behavioral} cognitive discounting parameter. In the robust confidence sets for the complete system the cognitive discounting \(\bar m\) is between 0.013 and 0.645 and in the robust confidence set for the single-equation estimations \(\bar{m}\) most of the time bellow one with at least 95\% coverage in the robust confidence sets and with a small trade-off between uncertainty and more tightness in the confidence sets. We are also able to take a novel approach where we do not pretest or screen for instruments prior to estimation. By reporting \(CS_R\), \(CS_N\) and \(\hat\Gamma\) we are able to provide the reader all ingredients one needs to interpret the results according to how much uncertainty one is willing to accept in exchange for tighter confidence sets. 

Taken together, these results have important implications for New Keynesian models, while still containing a certain degree of uncertainty as to where the cognitive discounting parameter lies, it seems clear that the parameter exist and could be an important ingredient for behavioral models.

\clearpage

\section{Tables and figures}
\label{sec:orgf747d60}

\begin{table}[!htbp]
\caption{\label{org9ed47be}
Maximum likelihood estimates of the complete behavioral DSGE model}
\begin{tabular}{@{}llll@{}}
\toprule
Parameters   & Estimate & s.d.   & t-stat  \\ \midrule
$\bar{m}$    & 0.6799   & 0.0704 & 9.6565  \\
$\gamma$     & 1.9709   & 0.4620 & 4.2662  \\
$\phi_\pi$   & 1.5058   & 0.2370 & 6.3543  \\
$\phi_x$     & 1.9672   & 0.2292 & 8.5844  \\
$\rho_i$     & 0.4623   & 0.0659 & 7.0120  \\
$\rho_d$     & 0.9591   & 0.0233 & 41.1592 \\
$\rho_m$     & 0.8843   & 0.0250 & 35.4231 \\
$\sigma^2_d$ & 0.6536   & 0.0946 & 6.9104  \\
$\sigma^2_s$ & 0.7443   & 0.0358 & 20.7978 \\
$\sigma^2_m$ & 1.0000   & 0.0000 & 0.0000  \\ \bottomrule
\end{tabular}
\end{table}

\begin{table}[!htbp]
\caption{\label{orgc0f1c14}
\(95\%\) \(LM_o\) confidence intervales parameters based on a single draw of simulated data and \(10^4\) draws over the parameter space. Notes: \(\sigma_{d,s,m}\) under construction.}
\centering
\begin{tabular}{@{}lccccccc@{}}
\toprule
Level & $\bar{m}$ & $\gamma$ & $\phi_\pi$ & $\phi_x$ & $\rho_i$ & $\rho_d$ & $\rho_m$ \\ \midrule
Lower & 0.013 & 1.82 & 0.70 & 0.51 & 0.22 & 0.90 & 0.79 \\
Upper & 0.645 & 4.91 & 0.98 & 1.55 & 0.44 & 0.98 & 0.98 \\ \bottomrule
\end{tabular}
\end{table}

\begin{table}[!htbp]
\caption{\label{orgb5f0b9c}
Confidence sets and distortion cutoffs \(\hat\Gamma\) for parameters myopia \(\bar{m}\) and risk aversion \(\gamma\) for the behavioral IS curve. Notes: \(\alpha=0.05\) and \(\Gamma_min=0.05\).}
\begin{tabular}{@{}cccc@{}}
\toprule
Parameter      & $CS_R$                                                           & $CS_N$           & $\hat\Gamma$ \\ \midrule
$\bar{m}$ & {[}0.00, 1.00{]}  & {[}0.80, 1.00{]} & 5.0\%          \\
$\gamma$       & {[}0.27, 10.00{]}  & {[}1.07, 3.49{]} & 5.0\%          \\ \bottomrule
\end{tabular}
\end{table}

\begin{table}[!htbp] \centering 
\caption{\label{org6b6ad4f}
Confidence sets and distortion Cutoffs \(\hat\Gamma\) for parameters myopia \(\bar{m}\) and risk aversion \(\gamma\) for the behavioral NKPC curve. Notes: \(\alpha=0.05\) and \(\Gamma_min=0.05\).}
\begin{tabular}{@{}cccc@{}}
\toprule
Parameter      & $CS_R$                                                           & $CS_N$           & $\hat\Gamma$ \\ \midrule
$\bar{m}$ & {[}0.07, 0.95{]}  & {[}0.14, 0.84{]} & 9.934\%          \\
$\gamma$       & {[}0.00, 10.00{]}  & {[}1.59, 10.00{]} & 9.934\%          \\ \bottomrule
\end{tabular}
\end{table}

\clearpage

\begin{figure}[htbp]
\centering
\includegraphics[width=\textwidth]{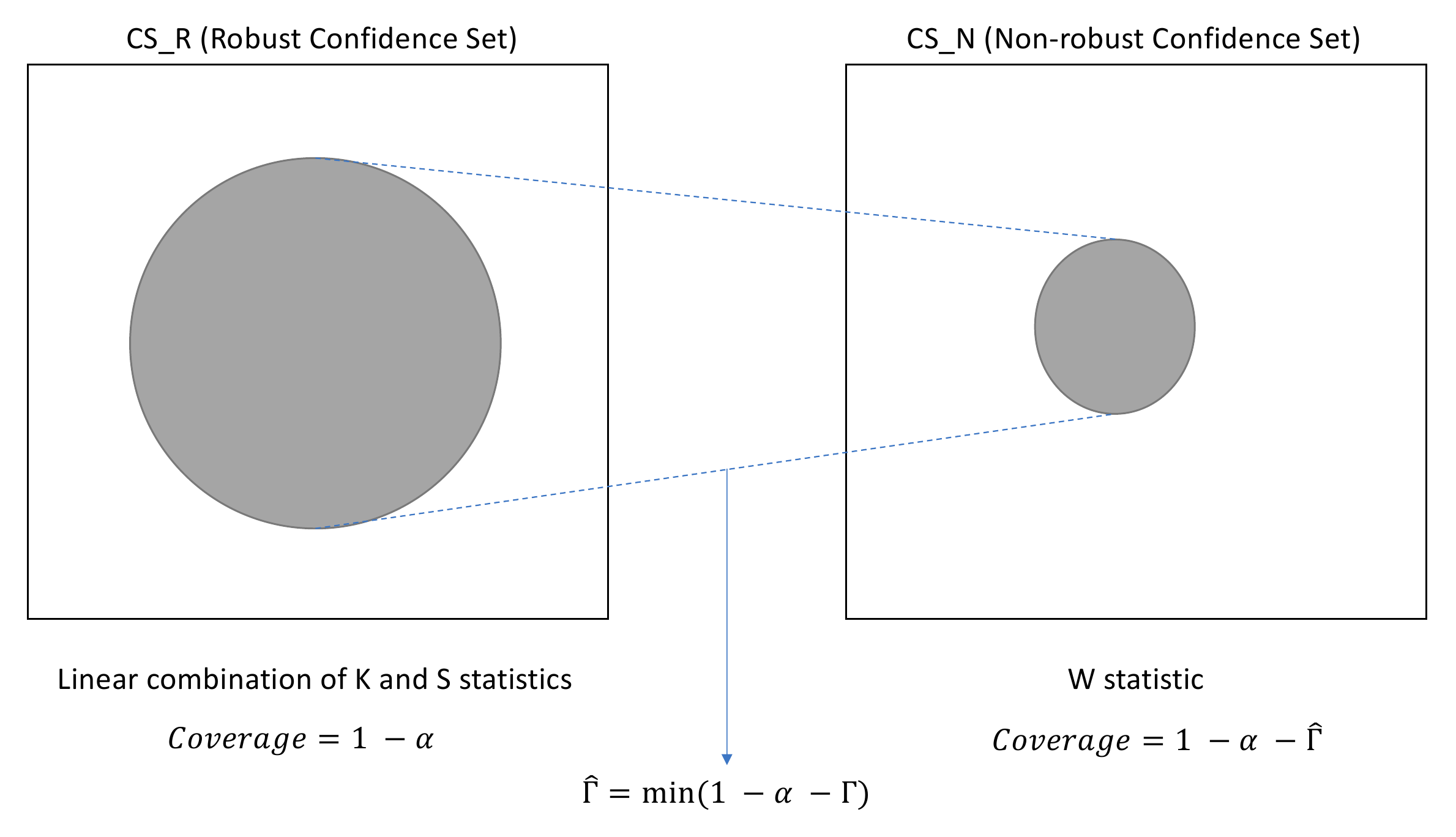}
\caption{\label{org16a9b50}
Illustration of the test procedure. Notes: \(\alpha\) is the desired coverage, usually \(5\%\) or \(10\%\) and \(\hat{\Gamma}\) is the minimal additional distortion that can be accepted to match the robust and non-robust sets.}
\end{figure}

\begin{figure}[htbp]
\centering
\includegraphics[width=\textwidth]{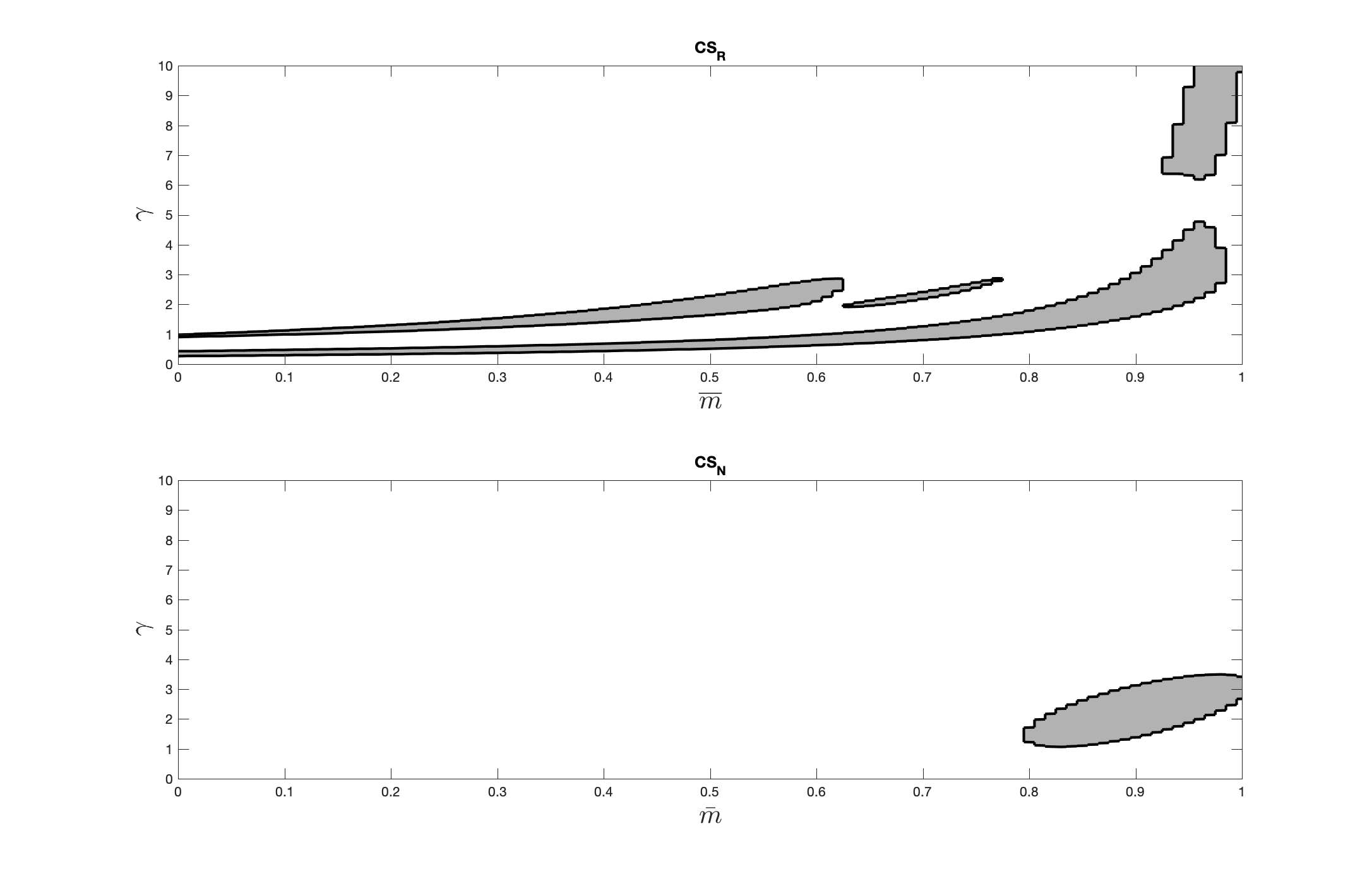}
\caption{\label{orgde53aaf}
Estimation of the behavioral IS curve using as instruments a constant and three lags of output gap and \((i_t - \pi_{t+1} - r_t)\) as in  \cite{ascarietal2016}. The CUGMM point estimates are \(\bar{m}=0.903\) and \(\gamma = 2.281\). The size of the test is \(\alpha = 0.05\) and the distortion cutoff \(\hat \Gamma\) is \(0.068\) for the entire set.}
\end{figure}

\begin{figure}[htbp]
\centering
\includegraphics[width=\textwidth]{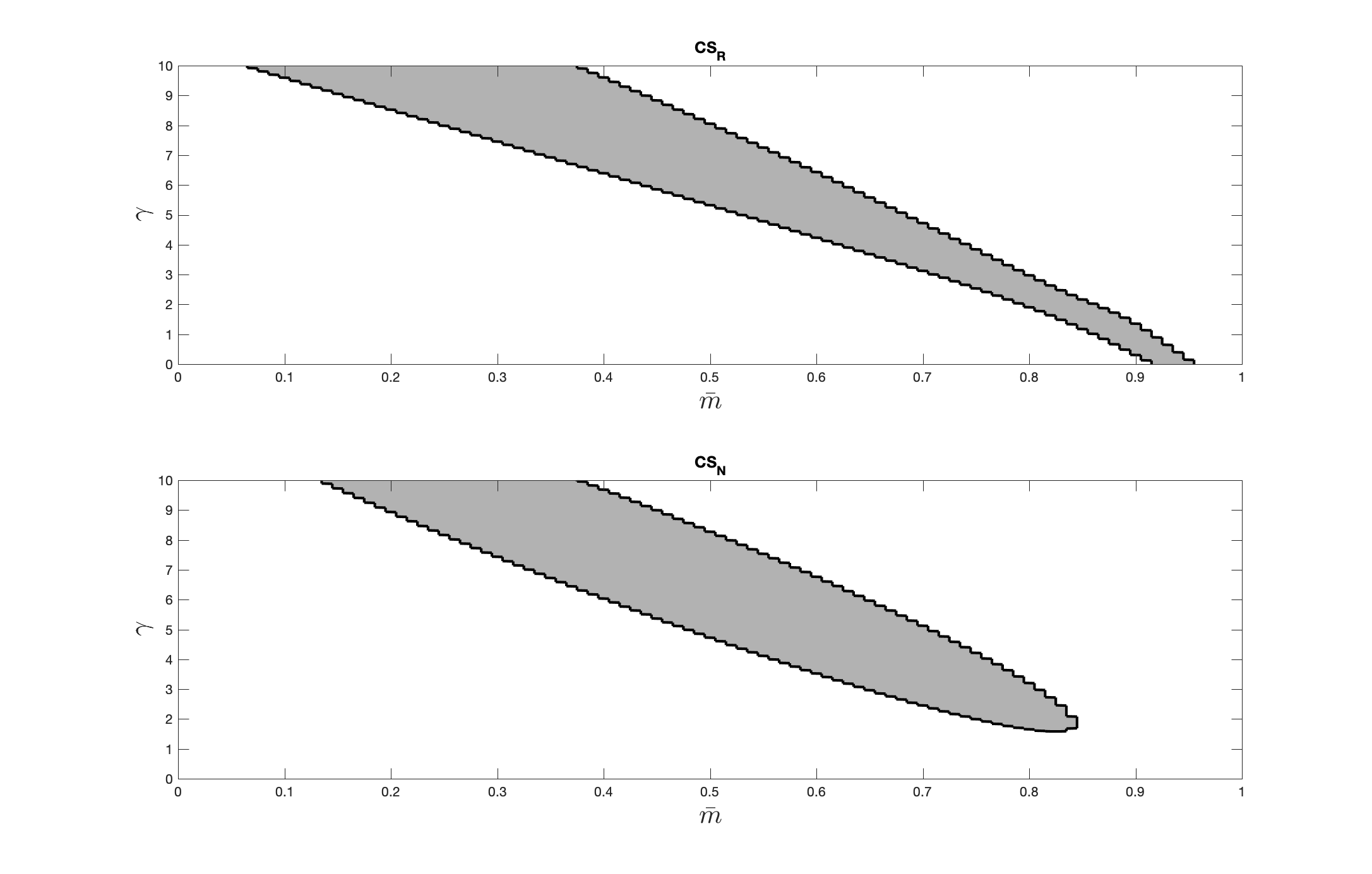}
\caption{\label{orgbeacf7f}
Estimation of the behavioral NKPC using as instruments four lags of inflation and three lags of the labor share, as in \cite{mavroeidisetal2014}. The CUGMM point estimates are \(\bar{m}=0.393\) and \(\gamma=7.944\). The size of the test is \(\alpha=0.05\) and the distortion cutoff \(\hat \Gamma\) is \(0.14\) for the entire set.}
\end{figure}

\clearpage

\section{Appendix}
\label{sec:orgd4e6a3c}
\subsection{DSGE solution}
\label{sec:org0af572b}

This section solves the DSGE model presented in Section \ref{sec:orgbf7538d}

\begin{equation}
\label{eq:org68a243c}
\begin{array}{rcl}
x_t & = & \bar{m}\mathbb{E}_t[x_{t+1}] - \sigma (i_t - \mathbb{E}_t\pi_{t+1}) + \eta_{d,t} \\
\pi_t & = & \beta M^f \mathbb{E}_t[\pi_{t+1}] + \kappa x_t + \epsilon_{s,t}\\
i_t & = & \rho_i i_{t-1} + (1 - \rho_i)(\phi_\pi \pi_t + \phi_x x_t) + \eta_{m,t}
\end{array}
\end{equation}
where the unobserved exogenous shocks evolve according to
\begin{equation*}
\begin{array}{rcl}
\eta_{d,t} & = & \rho_d \eta_{d,t-1} + \epsilon_{d,t}\\
\eta_{m,t} & = & \rho_m \eta_{m,t-1} + \epsilon_{m,t}\\
(\epsilon_{s,t}, \epsilon_{d,t}, \epsilon_{m,t})' & \sim & \text{i.i.d.}N(0,\Sigma)\\
\Sigma & = & \text{diag}(\sigma^2_s, \sigma^2_d, \sigma^2_m).
\end{array}
\end{equation*}
This is a restricted linear rational expectations system, to solve it we substitute out \(i_t\) and solve the expectations forward. First, substitute out \(i_t\) in the first equation of \ref{eq:org68a243c} and rearrange the terms with expectations to the left-hand side to obtain the system

\begin{equation*}
\begin{array}{lcl}
\bar{m}\mathbb{E}_t x_{t+1} + \sigma \mathbb{E}_t \pi_{t+1} & = & x_t + \sigma \dfrac{1}{\beta M^f} + \sigma \eta_{m,t} - \eta_{d,t}\\ [2ex]
\beta M^f \mathbb{E}_t \pi_{t+1} & = & -\kappa x_t + \pi_t. [2ex]
\end{array}
\end{equation*}
Solve for \(\mathbb{E}_t x_{t+1}\) and get the expectation equation

\begin{equation*}
\bar{m} \mathbb{E}_t x_{t+1} = (\beta M^f+\sigma\kappa)x_t + \beta M^f \sigma \eta_{m,t} - \beta M^f \eta_{d,t},
\end{equation*}
which can be rewritten as

\begin{equation*}
x_t = \frac{\bar{m}}{\beta M^f+\sigma\kappa} \mathbb{E}_t x_{t+1} - \frac{\beta M^f \sigma}{\beta M^f+\sigma\kappa} \eta_{m,t} + \frac{\beta M^f}{\beta M^f  + \sigma \kappa} \eta_{d,t}.
\end{equation*}
Now this expectation equation can be solved by forward iteration, giving

\begin{equation*}
x_t = \sum_{j=0}^{\infty} \left(\frac{\bar{m}}{\beta M^f+\sigma\kappa}\right)^j \mathbb{E}_t \left[-\frac{\beta M^f \sigma}{\beta M^f+\sigma\kappa} \eta_{m,t+j} + \frac{\beta M^f}{\beta M^f+\sigma\kappa} \eta_{d,t+j} \right].
\end{equation*}
Note that \(\mathbb{E}_t \eta_{i,t+j} = \rho_i^j \eta_{i,t}\) for \(i=d,m\), thus resulting in

\begin{equation*}
\begin{array}{rcl}
x_t & = & - \dfrac{\beta M^f \sigma}{\beta M^f + \sigma \kappa} \dfrac{1}{1-\rho_m \dfrac{\bar{m}}{\beta M^f + \sigma \kappa}} \eta_{m,t} + \dfrac{\beta M^f}{\beta M^f + \sigma \kappa} \dfrac{1}{1-\rho_d \dfrac{\bar{m}}{\beta M^f + \sigma \kappa}} \eta_{d,t}\\
{} & = &  - \dfrac{\beta M^f}{\beta M^f+\sigma\kappa -\bar{m}\rho_m} \eta_{m,t} + \dfrac{\beta M^f}{\beta M^f+\sigma\kappa - \bar{m}\rho_d} \eta_{d,t}.
\end{array}
\end{equation*}
Substitute the last expression into the IS equation and repeat the same process solving the resulting expectation equation for \(\pi_t\)

\begin{equation*}
\begin{array}{rcl}
\pi_t & = & \beta M^f \mathbb{E}_t \pi_{t+1} + \kappa x_t \\ [2ex]
{} & = & \beta M^f \mathbb{E}_t \pi_{t+1} - \dfrac{\beta M^f \sigma \kappa}{\beta M^f + \sigma \kappa - \bar{m} \rho_m} \eta_{m,t} + \dfrac{\beta M^f \kappa}{\beta M^f + \sigma \kappa - \bar{m} \rho_d} \eta_{d,t} \\ [2ex]
{} & = & \sum_{j=0}^{\infty} (\beta M^f)^j \mathbb{E}_t \left[-\dfrac{\beta M^f \sigma \kappa}{\beta M^f+\sigma\kappa - \bar{m} \rho_m} \eta_{m,t+j} + \dfrac{\beta M^f \kappa}{\beta M^f+\sigma\kappa -\bar{m} \rho_d} \eta_{d,t+j} \right] \\ [2ex]
{} & = & - \dfrac{\beta M^f \sigma \kappa}{(\beta M^f + \sigma \kappa - \bar{m} \rho_m)(1 - \rho_m \beta M^f)} \eta_{m,t} + \dfrac{\beta M^f \kappa}{(\beta M^f + \sigma \kappa - \bar{m} \rho_d)(1 - \rho_d \beta M^f)} \eta_{d,t}.
\end{array}
\end{equation*}

One obtains, therefore, the solution to the system \ref{eq:org68a243c}:

\begin{equation*}
\begin{array}{rcl}
x_t & = &  - \dfrac{\beta M^f}{\beta M^f+\sigma\kappa -\bar{m}\rho_m} \eta_{m,t} + \dfrac{\beta M^f}{\beta M^f+\sigma\kappa - \bar{m}\rho_d} \eta_{d,t}\\[2ex]
\pi_t & = & - \dfrac{\beta M^f \sigma \kappa}{(\beta M^f + \sigma \kappa - \bar{m} \rho_m)(1 - \rho_m \beta M^f)} \eta_{m,t} + \dfrac{\beta M^f \kappa}{(\beta M^f + \sigma \kappa - \bar{m} \rho_d)(1 - \rho_d \beta M^f)} \eta_{d,t}.
\end{array}
\end{equation*}

\subsection{Two-step identification-robust confidence sets algorithm}
\label{sec:org1a4b023}

This section details the test and is entirely based on \textcite{andrews2018}. In GMM models, for all the commonly-used non-robust confidence sets \(CS_{N}\) and any \(\Gamma > 0\) we can construct preliminary robust confidence set \(CS_{P}\) with the same coverage regardless of identification strength and which is contained in the non-robust set with probability one under strong identification. For this we define the \(S\) statistic of \cite{stockwright2000} and K statistic of \cite{kleibergen2005}. The problem is the S statistics is inefficient with over identification and the \(K\) statistic is often inconsistent (\textit{i.e.} fails to shrink towards the true parameter even as the sample grows because it gathers local minima and maxima) with the equivalency with the Wald confidence set holding only locally, not globally.

Thus, to obtain a consistent confidence set, for \(a>0\) consider \(CS_R = ( \vartheta : K(\vartheta) + a . S(\vartheta) \leq \chi^2_{1,1-\alpha})\) where \(K(\vartheta) + a . S(\vartheta)\) is a linear combination statistic, as in \cite{andrews2016}. This confidence set has coverage \(1 - \alpha - \Gamma(a) = Pr((1+a). \chi^2_1 + a .  \chi^2_{k-1} \leq  \chi^2_{1,1-\alpha} )\) regardless of identification strength. So  \(\Gamma \to 0\) as \(a \to 0\), and we can choose \(a\) to obtain any desired level of \(\Gamma\).

Now pick some \(\Gamma_{min} \geq 0\). For \(\Gamma \geq \Gamma_{min}\), consider the family of robust confidence sets \(CS_{P} = ( \vartheta : K(\vartheta) + a(\Gamma) . S(\vartheta) \leq \chi^2_{1,1-\alpha})\) where \(CS_{P} (\Gamma) = (\vartheta : K_{\Gamma}(\vartheta) \leq \chi^2_{k,1-\alpha} )\) is designed to have coverage exceeding \(1 - \alpha - \Gamma\).

Define the robust confidence set as \(CS_R(\Gamma) = (\vartheta : K_{\Gamma}(\vartheta) \leq H^{-1}_{k,1-\alpha})\), where \(H_{k,1-\alpha}\) is \(1-\alpha\) quantile of \((1 + a(\Gamma)\chi^2_1 + a(\Gamma)\chi^2_{k-1})\) and has the correct critical values with coverage exceeding \(1-\alpha\).\footnote{\(H(x;a,k,p)\) is the cumulative distribution function for the a \((1+a) \times \chi^2_p + a \times \chi^2_{k-p}\) distribution, which is a linear combination of \(\chi^2\) variables, and \(H^{-1}(1-\alpha;a,k,p)\) the \(1-\alpha\) quantile of this distribution} And we end up with the following two-step confidence set

\begin{equation}
  CS_{2S}(\Gamma)=\begin{cases}
    CS_{N} & \text{if $CS_{P}(\Gamma) \subseteq CS_{N}$}\\
    CS_R(\Gamma) & \text{if $CS_{P}(\Gamma) \nsubseteq CS_{N}$.}
  \end{cases}
\end{equation}

Note that these preliminary confidence sets are decreasing in \(\Gamma\): \(\Gamma \leq \Gamma' \implies CS_{P}(\Gamma') \subseteq CS_{P}(\Gamma)\). Thus, we have the property that \(CS_{P}(\Gamma) \subseteq CS_2(\Gamma)\), so \(CS_2(\gamma)\) has coverage exceeding \(CS_{P}(\Gamma)\) which exceeds \(1 - \alpha -\Gamma\). Therefore, we get a bounded size distortion. Also note that under strong identification we have \(CS_{P}(\Gamma) \subseteq CS_{N}\) so \(CS_2(\Gamma) = CS_{N}\) asymptotically. Now define the maximal distortion cutoff as \(\hat\Gamma = min (\Gamma \geq \Gamma_{min} : CS_{P} (\Gamma) \subseteq CS_{NR})\) and report \(CS_{N}\), \(CS_R(\hat\Gamma)\), and \(\hat\Gamma\).

To empirically implement \(CR_R\), \(CS_N\), we set \(\Gamma_{min}\) equal to \(5\%\) and \(\alpha\) equal to \(10\%\) so coverage of the robust set is at least \(90\%\) and apply \cite{andrews2018} six-step algorithm:

\begin{enumerate}
\item \textit{Choose the weighting matrix and estimator}. As already mentioned, we use the CUGMM of the form of equation \ref{eq:org8a782df} with \(\hat{W}(\vartheta) = \hat\Sigma(\vartheta)^{-1}\) as the efficient weighting matrix. Then define the Wald statistic, where \(\hat\Sigma_{\hat\beta}\) is the usual GMM variance estimator for \(f(\hat\vartheta)\).

\item \textit{Choose grid of parameter values}. Since to calculate the confidence sets we work with test inversions we need to discretize the parameter space to obtain all values where the test statistics falls bellow given thresholds. In this implementation we consider

$$\vartheta^1 = (\bar{m}, \gamma) \in \Theta^1_D = (0,0.1,...,1) \times (0,0.1,...,10)$$

and

$$\vartheta^2 = (\bar{m}, \gamma) \in \Theta^2_D = (0,0.1,...,1) \times (0,0.1,...,10).$$

Let \(\Theta_D\) represent the elements of \(\Theta^1_D\) and \(\Theta^2_D\), which are \((\vartheta^i_1,...,\vartheta^1_{|\Theta_D|}) \forall i=1,2\).

\item \textit{Calculate test statistics}. Given this discrete approximation to the parameter space, for each \(\vartheta^i_n \in \Theta_D\) we can calculate \(S_T(\vartheta^i_n)\) and \(\hat\Sigma(\vartheta^i_n\) and the test statistics \(S\), \(K\) and \(W\).

\item \textit{Calculate} \(a(\Gamma_{min})\). Now, determine the value of \(a(\Gamma_{min})\) to be used in the construction of the robust confidence set.

\item \textit{Calculate} \(CR_R\) and \(CS_N\). With \(a(\Gamma_{min})\) we can calculate the critical value used in \(H^{-1}\). The robust confidence is then

$$CS_R = (f(\vartheta^i_n): \vartheta^i_n \in \Theta_D, K_{\sigma,f}(\vartheta^i_n) + a \times S(\vartheta^i_n) \leq H^{-1}(1-\alpha, a(\Gamma_{min}),k,p))$$

and the nonrobust confidence set is

$$CS_N = (f(\vartheta^i_n): \vartheta^i_n \in \Theta_D, W(f(\vartheta^i_n)) \leq \chi^2_{p,1-\alpha}).$$

\item \textit{Calculate} \(\hat\Gamma\). Finally, the distortion cutoff can be calculated.
\end{enumerate}

\subsection{Two-step confidence sets for \(\alpha=0.1\)}
\label{sec:orgbc247d4}

For the behavioral IS curve the parameter grid used was \(\vartheta = (\bar{m}, \gamma) \in \Theta_D = {0.01, 0.02, ..., 0.99} \times {0.01, 0.02, ..., 5}\), with \(\alpha = 0.1\) and \(\Gamma_min = 0.05\). Table \ref{orgb5f0b9c} presents the results illustrated in Figure \ref{orgde53aaf}. The CUGMM point estimates are \(\bar{m}=0.9029\) and \(\gamma = 2.281\) and the distortion cutoff \(\hat \Gamma\) is \(0.08\) for the entire set. This means that for one to believe in the non-robust set one has to be willing to add \(8\%\) on top of the original size of \(5\%\). Looking individually the distortion cutoff are a bit lower at \(0.053\) for \(\bar m\) and \(0.052\) for \(\gamma\). The robust set is valid at the \(5\%\) level. However, in this case if we consider the controlled size distortion we can then have a bounded set for both variables. If one is willing to make the uncertainty trade-off, then the predicted value of \(\bar m\) lies between \(0.81\) and \(0.99\), while for \(\gamma\) it is between \(1.22\) and \(3.34\).

Figure \ref{orgbeacf7f} illustrates the results for the behavioral NKPC estimation. The parameter grid was the same with \(\alpha = 0.1\) and \(\Gamma_min = 0.05\) as well. The CUGMM point estimates are \(\bar{m}=0.393\) and \(\gamma=7.944\) and the distortion cutoff \(\hat \Gamma\) is \(0.16\) for the entire set. Table \ref{org6b6ad4f} details the results for each parameter. The distortion cutoff is a bit lower at \(11.16\%\) each. More importantly, there is an upper and lower bound for \(\bar m\).

\begin{table}[!htbp]
\caption{\label{orgb625390}
Confidence Sets and Distortion Cutoffs \(\hat\Gamma\) for parameters \(\bar{m}\) and \(\gamma\) for the behavioral IS curves. Notes: \(\alpha=0.10\) and \(\Gamma_min=0.05\).}
\begin{tabular}{@{}cccc@{}}
\toprule
Parameter      & $CS_R$                                                           & $CS_N$           & $\hat\Gamma$ \\ \midrule
$\bar{m}$ & {[}0.01, 0.97{]}  & {[}0.81, 0.99{]} & 5.3\%          \\
$\gamma$       & {[}0.28, 4.31{]}  & {[}1.22, 3.34{]} & 5.2\%          \\ \bottomrule
\end{tabular}
\end{table}

\begin{table}[!htbp] \centering 
\caption{\label{org51427f0}
Confidence Sets and Distortion Cutoffs \(\hat\Gamma\) for parameters \(\bar{m}\) and \(\gamma\) for the behavioral NKPC curve. Notes: \(\alpha=0.10\) and \(\Gamma_min=0.05\).}
\begin{tabular}{@{}cccc@{}}
\toprule
Parameter      & $CS_R$                                                           & $CS_N$           & $\hat\Gamma$ \\ \midrule
$\bar{m}$ & {[}0.55, 0.95{]}  & {[}0.50, 0.78{]} & 11.16\%          \\
$\gamma$       & {[}0, 5.0{]}  & {[}2.37, 5.00{]} & 11.16\%          \\ \bottomrule
\end{tabular}
\end{table}

\begin{figure}[htbp]
\centering
\includegraphics[width=\textwidth]{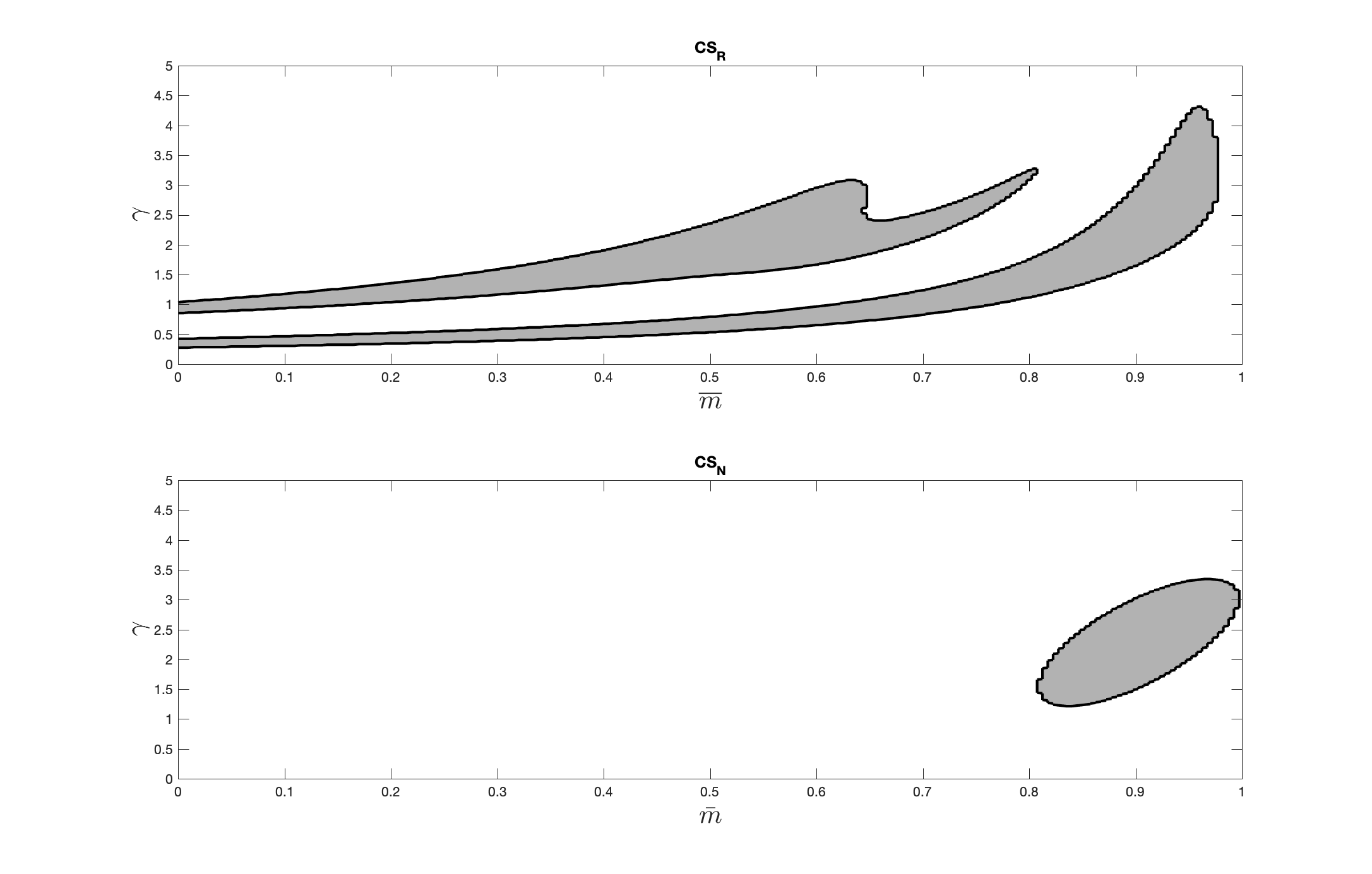}
\caption{\label{org50f6158}
Estimation of the behavioral IS curve using as instruments a constant and three lags of output gap and \((i_t - \pi_{t+1} - r_t)\) as in  \cite{ascarietal2016}. The CUGMM point estimates are \(\bar{m}=0.903\) and \(\gamma = 2.281\). The size of the test is \(\alpha = 0.1\) and the distortion cutoff \(\hat \Gamma\) is \(0.08\) for the entire set.}
\end{figure}

\begin{figure}[htbp]
\centering
\includegraphics[width=\textwidth]{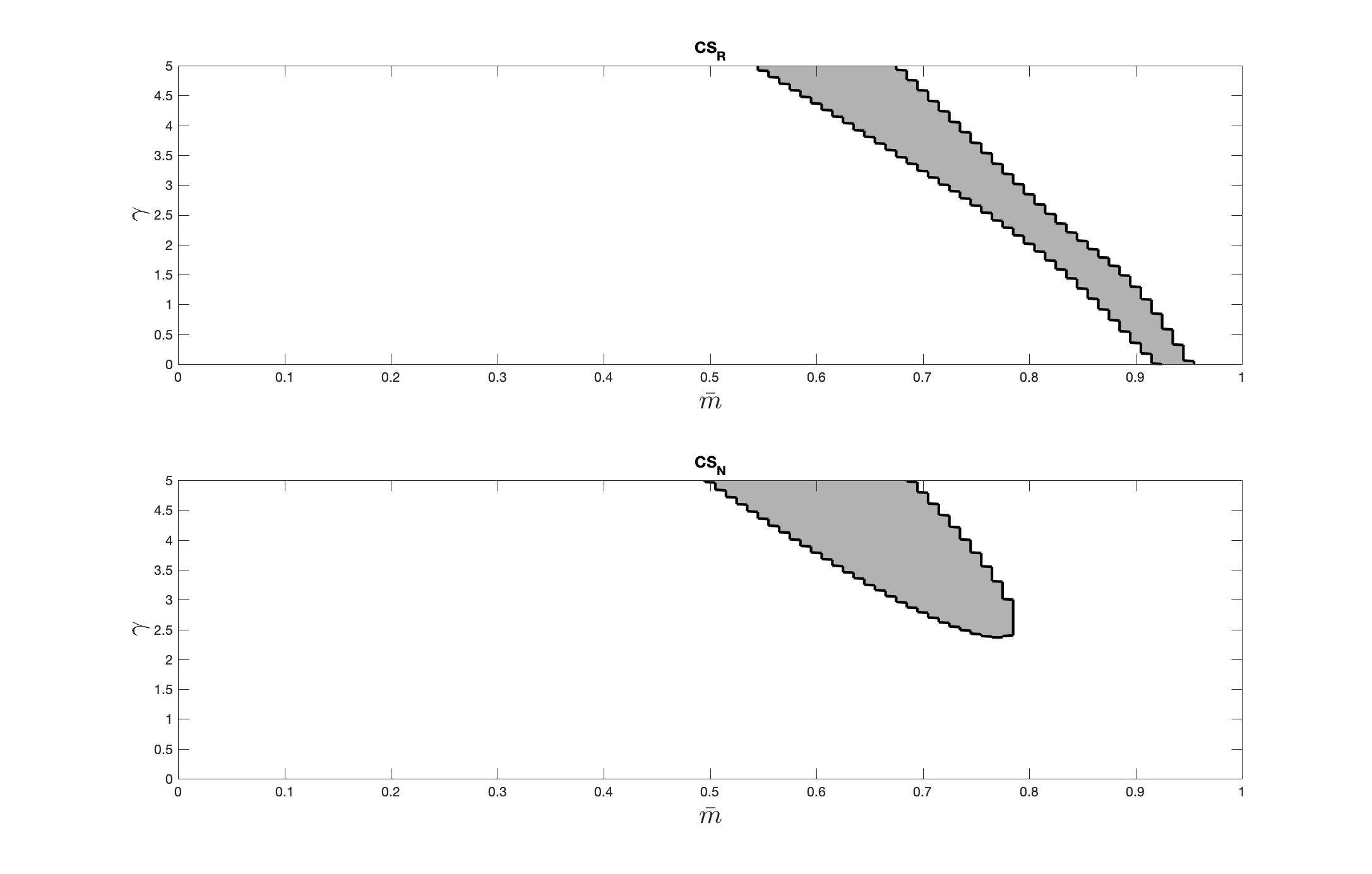}
\caption{\label{org49d4881}
Estimation of the behavioral NKPC using as instruments four lags of inflation and three lags of the labor share, as in \cite{mavroeidisetal2014}. The CUGMM point estimates are \(\bar{m}=0.393\) and \(\gamma=7.944\). The size of the test is \(\alpha=0.1\) and the distortion cutoff \(\hat \Gamma\) is \(0.16\) for the entire set.}
\end{figure}

\clearpage

\printbibliography
\end{document}